# Diffuse X-ray emission around high-redshift, radio-loud QSOs

Matthias Bartelmann[1], Peter Schneider[1], and Günther Hasinger[2]

[1]Max-Planck-Institut für Astrophysik, Postfach 1523, D–85740 Garching, FRG; [2]Max-Planck-Institut für extraterrestrische Physik, Gießenbachstraße 2, D–85740 Garching, FRG

**Abstract.** We announce the detection of correlations on angular scales of $\gtrsim 10'$ between optically bright, high-redshift, radio-loud QSOs with diffuse X-ray emission seen by ROSAT in the *All-Sky Survey*. These correlations reach significance levels of up to 99.8%. A comparison of the results with a sample of control fields, bootstrapping analyses, and Kolmogorov-Smirnov tests provide unambiguous support for the statistical significance of the correlations found. We argue that the detected enhanced diffuse X-ray emission is in the foreground of the QSOs, and that it is probably due to galaxy clusters which magnify the QSOs by their gravitational lensing effect, thereby giving rise to a magnification bias in the background source sample. A comparison of the results presented below with correlations previously found between the same QSO sample and either Lick or IRAS galaxies provides further evidence for this interpretation, and identifies positions in the sky where weak gravitational lensing may be detected by searching for coherent distortions of background galaxy images.

## 1 Introduction

Previous analyses of correlations between foreground galaxies and distant, optically bright, radio-loud QSOs on angular scales of $\gtrsim 10'$ (hereafter called large-scale correlations) have shown that these QSOs are correlated with galaxies selected by either infrared or optical emission (Bartelmann & Schneider 1993b,1994; henceforth BS2,3). Trends in the results as well as their qualitative agreement with a previously performed theoretical investigation (Bartelmann & Schneider 1993a, henceforth BS1) provide evidence that these correlations are caused by the magnification bias due to weak lensing by extended inhomogeneities in the Universe, on scales of galaxy clusters or larger.

Clusters, probably consisting mainly of dark matter, are usually X-ray bright objects, with luminosities in the range $L_{\rm X} \in [10^{43} \ldots 10^{45}]$ ergs/s (e.g., Sarazin 1992 and references therein). Therefore, if clusters are responsible for the large-scale correlations in question, one should be able to see enhanced diffuse X-ray emission in the vicinity of bright and distant, presumably weakly lensed background sources. This paper describes the results of a correlation analysis on the basis of data from the ROSAT *All-Sky Survey* (see, e.g., Voges 1992).

We employ the rank-order correlation test described and applied in BS1,2,3, which has proven to be highly sensitive to detect weak correlations. The advantage of the rank-order correlation test is that it is a non-parametric, distribution-free test which is unaffected by linear transformations of the data sets tested (and therefore unaffected by normalization problems); its disadvantage is that it yields a 'binary' result, since it merely states whether there are significant correlations or not.

The paper is organized in the following way. Sect.2 briefly describes the selection of sources and of control fields. Sect.3 gives the average X-ray photon number densities retrieved from the ROSAT *All-Sky Survey* in those fields. Sect.4 presents the rank-order correlation results and a bootstrapping test of their reliability, and Sect.5 describes Kolmogorov-Smirnov tests further supporting the results of Sect.4. Sect.6 provides an explanation of a surprising feature detected in the analysis of the control fields. In Sect.7, we introduce further statistical tests with the focus on illustrating the rank-order correlation results. Sect.8 is devoted to a search for such background sources which either are close to known X-ray bright galaxy clusters or which appear significantly correlated with X-ray emission *and* galaxies. Finally, Sect.9 summarizes and discusses the results.

## 2 Selection of sources and control fields

We use the same background source sample as in BS2,3, namely the optically identified 1-Jansky sources (Kühr et al. 1981, Stickel 1992, Stickel & Kühr 1993a,b) with redshifts above 0.5; see BS2,3 for a more detailed description of this sample. The sample contains 246 sources, most of which are flat-spectrum QSOs. The numbers and positions of X-ray photons in the ROSAT energy range $E_\gamma \in [0.1, 2.2]$ keV in fields of $2° \times 2°$ size centered on these sources were obtained from the *All-Sky Survey*, together with an equal number of equally sized control fields with the same galactic latitude as the source fields, but with randomly chosen galactic longitudes. The control fields were chosen such that the contamination by galactic foreground emission should on average be equal in the source and in the control fields.

## 3 Mean photon counts

In order to avoid severe contamination of the results by anisotropic X-ray noise, we restrict the analysis to photons with energies $\geq 0.75$ keV. The anisotropic noise in the *All-Sky Survey* is primarily due to line emission of atmospheric oxygen at 0.54 keV. Since it is therefore mainly caused by the Earth's atmosphere, it depends on the attitude of the ROSAT telescope relative to the Earth and the Sun, resulting in a pattern of stripes irregularly covering the X-ray sky as seen in the *All-Sky Survey*. Above $\simeq 0.75$ keV, these stripes dim and disappear such that the X-ray sky is basically unaffected by atmospheric (and therefore attitude-dependent) X-ray noise in the energy ranges which we consider here (Freyberg 1993). We distinguish between three different energy ranges of the X-ray photons, namely $E_\gamma \geq 0.75$ keV, $E_\gamma \geq 1.0$ keV, and $E_\gamma \geq 1.5$ keV (later on called 'soft', 'medium', and 'hard band', respectively).

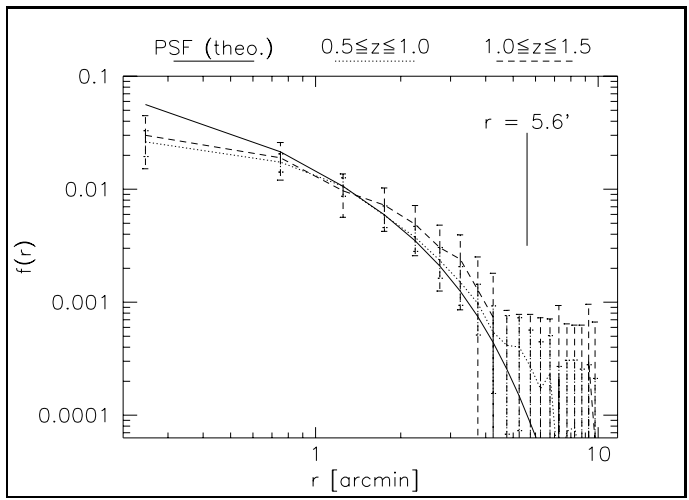

**Fig. 1.** Radial photon-count profiles obtained from stacks of source fields, compared with the theoretical point-spread function for the *All-Sky Survey*. The three curves are the theoretical PSF (solid line) and the radial profiles of the QSO X-ray emission for redshifts $z \in [1.0, 1.5]$ and $z \in [1.5, 2.0]$ (dotted and dashed curves, respectively); only photons with $E_\gamma \geq 0.75$ keV were selected. The background is subtracted and the curves are normalized to unity within $10'$. The error bars on the dashed curve are Poisson errors

On average, there is only weak emission in the source as well as in the control fields. The mean photon counts in source and control fields are statistically identical, irrespective of the energy band. We obtain an average photon density $\langle n_\gamma \rangle \simeq 0.19, 0.12, 0.05\ (')^{-2}$ in the soft, medium, and hard regimes, respectively.

The photon number density in a field depends, of course, also on the exposure time of this field. Therefore, it is enhanced in fields close to the ecliptic poles, which are covered by ROSAT once per orbit. This does, however, not affect the later correlation analysis since the rank-order test is insensitive to linear transformations of the data.

To illustrate the point-spread function, we display in Fig.1 the theoretical point-spread function for the *All-Sky Survey* (solid line; cf. Hasinger et al. 1992) and two radial photon-count profiles obtained from stacks of source fields for $E_\gamma \geq 0.75$ keV. These two curves are for QSOs within redshift bins $z \in [0.5, 1.0]$ and $z \in [1.0, 1.5]$ (dotted and dashed curves, respectively). The background is subtracted, and the curves are normalized to unity within $10'$. The point-spread function is sufficiently narrow to ensure that outside $r \gtrsim 5'.6$ [1] there is only negligible contamination from the central sources themselves. Within the error bars, the radial source profiles agree well with the theoretical PSF, except perhaps for the innermost data point. However, the source profiles were obtained from pixelized image stacks, which may cause a misalignment of the order of one pixel ($30''$) between pixel centers and the true source positions, affecting the height of the profile.

## 4 Rank-order correlation results

The rank-order correlation test is performed as described in BS1,2,3. Briefly, the method proceeds as follows. A field in the sky centered on the position of a (background) source is covered by a pattern of $N_{\rm cell}$ (we choose $N_{\rm cell} = 25$) equal-area cells, which, where not stated otherwise, are chosen to be rings of area 100 square arcminutes concentric with the source position. To avoid contamination from the emission of the background sources themselves, we later cut out the central cell and restrict the correlation analysis to the $N'_{\rm cell} = 24$ surrounding cells, covering a ring between radii $5'.6$ and $28'.2$. The distances of these rings to the center are ranked in descending order, providing a set of distance ranks $\{\mathcal{R}_\mathrm{d}\}$. Then, the 'events' (galaxies in BS1,2,3 or X-ray photons in this paper) in these cells are counted and assigned count ranks $\mathcal{R}_\mathrm{c}$ in ascending order. The two sets of ranks per field, $\{\mathcal{R}_\mathrm{d}\}$ and $\{\mathcal{R}_\mathrm{c}\}$, define a hierarchy of 'closeness' to the sources and of counts, respectively. These two hierarchies are compared, and the comparison is quantified by a correlation coefficient $r_\mathrm{corr} \in [-1, 1]$. The statistical properties of $r_\mathrm{corr}$ are known, and therefore $r_\mathrm{corr}$ can be uniquely translated into an error level $\epsilon \in [0, 1]$. $\epsilon$ is the error probability for the correlation hypothesis, or, conversely, $(1-\epsilon)$ is the statistical significance of the correlation result. For further details on this correlation technique, the reader is referred to BS1,2,3 or Kendall & Stuart (1973).

---

[1] for the choice of this number, see the next section

The rank-order correlation test can be performed either with individual sources, yielding one correlation coefficient, and thus one error level per source, or with the averaged hierarchies of a source subsample, yielding the correlation error level of the whole subsample. In Fig.2, we show examples of source and control fields which were selected for either low or high correlation coefficients. As in BS2,3, we define source subsamples by three parameters, namely an optical (visual) magnitude threshold $m_\mathrm{max}$, controlling the optical brightness limit of the subsample, and the bounds of a redshift interval $[z_\mathrm{min}, z_\mathrm{max}]$, controlling the average distance of the subsample. Where $z_\mathrm{max}$ is not given, it is set to infinity.

Subsamples can equally be defined for the control fields, since there is one control field per source. We assign each control field the redshift and the optical magnitude of the corresponding source. In this case, the subsample parameters have the sole effect of reducing the field number by the same amount as in the source subsamples.

To estimate the error $\Delta\epsilon$ of $\epsilon$ obtained from either source or control fields, we perform a bootstrapping analysis. This is done by randomly selecting from the $N$ sources or control fields contained in the subsample $N$ fields, but allowing fields to repeatedly enter the bootstrapped sample (see, e.g., Press et al. 1992 and references therein). This yields a (cumulative) distribution $P(\epsilon)$ of error levels per subsample, with $P(100\%) = 1$, $P(0\%) = 0$. We call the median $\epsilon'$ and choose as an error estimate the range $\Delta\epsilon$ within which $P(\epsilon) \in [0.1, 0.9]$.

Fig.3 shows $\epsilon$, $\epsilon'$, and $\Delta\epsilon$ obtained from source and control fields for finite $z_\mathrm{max}$, Fig.4 shows the results for $z_\mathrm{max} = \infty$.

First of all, the figures clearly show evidence for a difference between the results obtained from source fields [left columns in Figs.(3,4)] and those obtained from control fields [right columns in Figs.(3,4)]. In particular, Fig.3 shows that

(1) visually bright ($m_\mathrm{max} \lesssim 19$), close ($0.5 \leq z \leq 1$) 1-Jansky sources are correlated with soft- and medium-energy X-ray photons,

(2) 1-Jansky sources in the intermediate redshift interval ($1.0 \leq z \leq 1.5$) do not show significant correlation with X-ray photons irrespective of their visual brightness and of the photon energy,

(3) distant 1-Jansky sources ($1.5 \leq z \leq 2.0$) are strongly correlated with soft and hard X-ray photons, and

(4) the results from the control fields show no indication for any significant correlation irrespective of the X-ray photon energy.

(5) The correlations show a trend to increase with increasing optical-brightness threshold of the QSO subsamples.

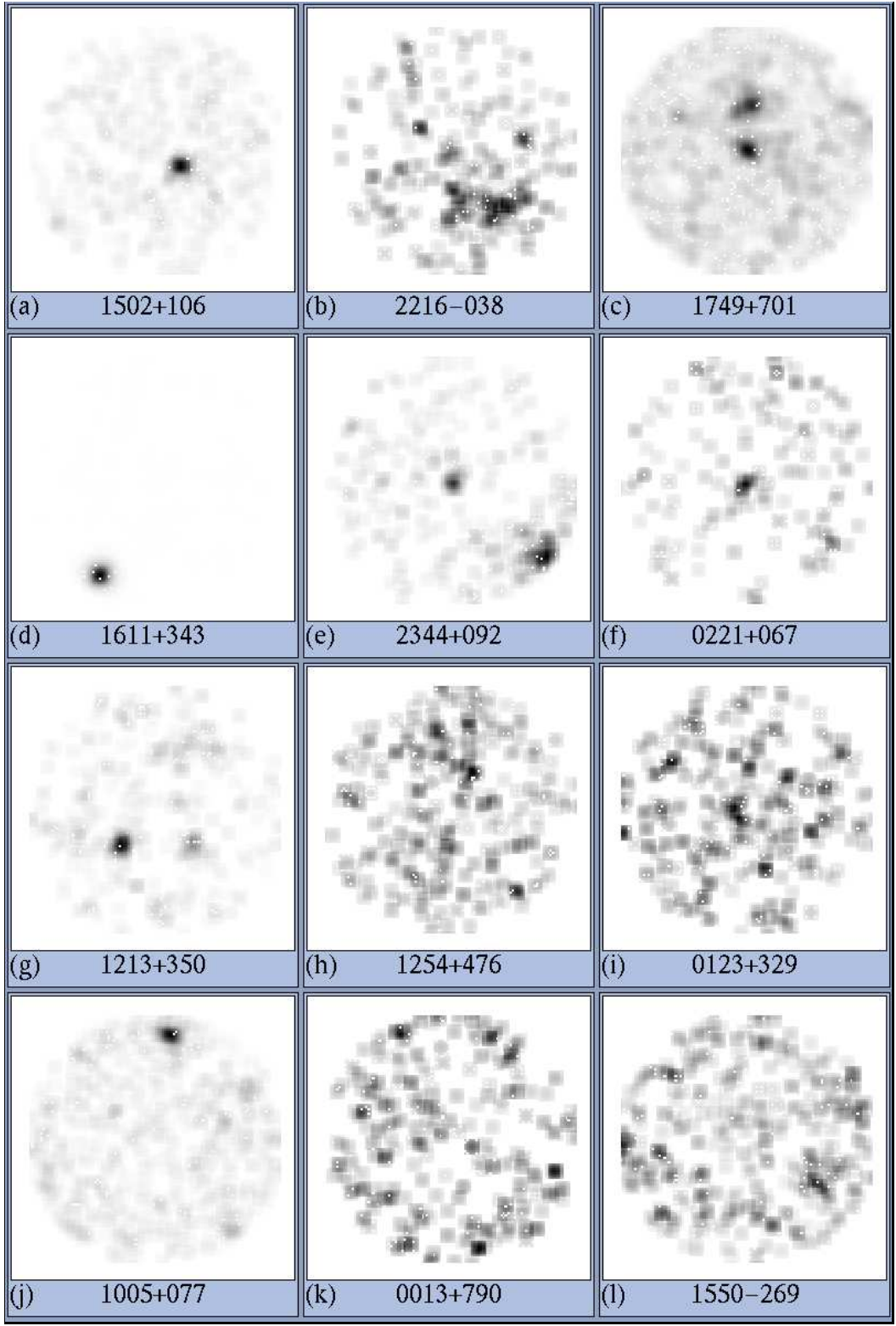

**Fig. 2** Examples of source and control fields. Panels (a) to (c): source fields with the highest $r_\mathrm{corr}$, panels (d) to (f): source fields with the lowest $r_\mathrm{corr}$; panels (g) to (l): the same for control fields. In most of the panels, the exposure time is sufficiently short so that individual photons can be seen. The white patches are an artefact from data reduction and smoothing with the EXSAS software package. The extended sources in panel (b) show two EMSS foreground clusters listed in Tab.6

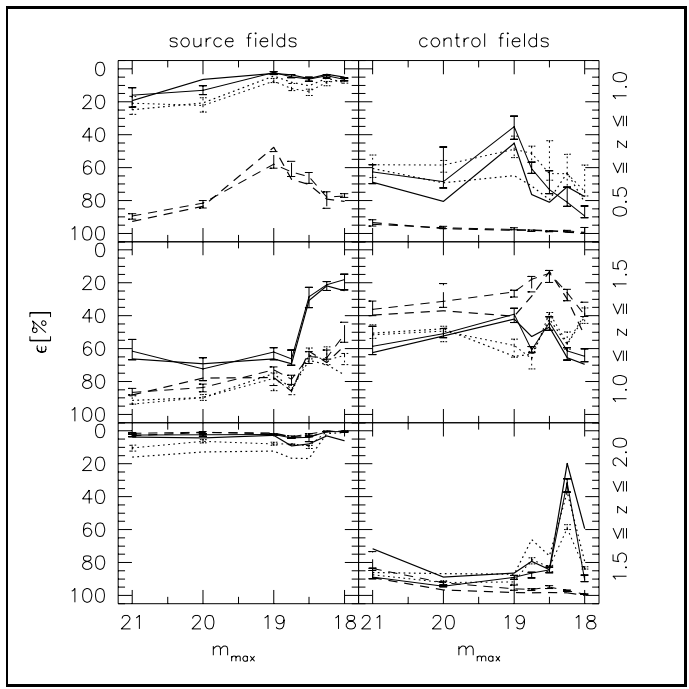

**Fig. 3.** Rank-order correlation results for source fields (left column) and control fields (right column). Field subsamples become brighter to the right (decreasing $m_{\max}$), correlations and anticorrelations become more significant to the top and the bottom of the plots (decreasing and increasing error level $\epsilon$), respectively. The thick lines show $\epsilon$ (the error level from the data), the thin lines $\epsilon'$ (the median error level from bootstrapped data samples), and the error bars show the range $\Delta\epsilon$ within which 80% of the bootstrapped data samples fall. The three rows show the results for different redshift intervals. Line types distinguish between the energy ranges $E_\gamma \geq \{0.75, 1.0, 1.5\}$ keV (solid, dotted, and dashed lines, respectively)

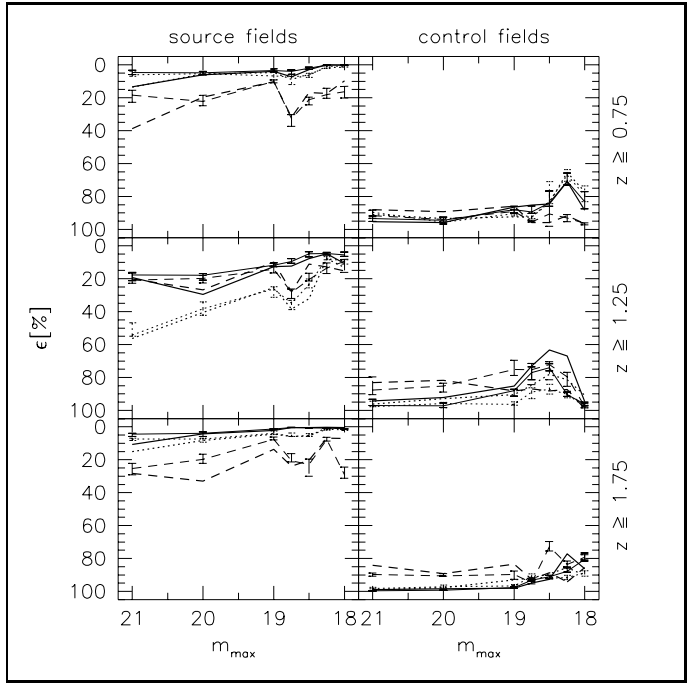

**Fig. 4.** Similar to Fig.3, but for $z_{\max} = \infty$; the panels in the three rows show the results for different lower redshift limits $z_{\min}$

For more detailed information, we have also listed the results in Tables 1,2, and the bootstrapping results in Tab.3.

The bootstrapping results confirm the results obtained from the original subsamples, indicating that the results are not dominated by a few sources only, and the errors $\Delta\epsilon$ are usually small compared to $\epsilon$ or $\epsilon'$.

It is surprising that no significant correlation of X-ray emission with intermediate-redshift 1-Jansky sources ($1.0 \leq z \leq 1.5$) is found, while closer as well as more distant

**Table 1.** Correlation results obtained from source and control fields (data displayed in Fig.3). $z_{\min,\max}$ are the limiting redshifts of the source subsamples, $m_{\max}$ is the maximum optical magnitude of the sources, and $N$ is their number in the subsample. $\epsilon_{\geq E}$ is the error level of the correlation in percent, where $E$ is the minimum photon energy in keV

| $z_{\min}$ | $z_{\max}$ | $m_{\max}$ | $N$ | source fields | | | control fields | | |
|---|---|---|---|---|---|---|---|---|---|
| | | | | $\epsilon_{>0.75}$ | $\epsilon_{>1.0}$ | $\epsilon_{>1.5}$ | $\epsilon_{>0.75}$ | $\epsilon_{>1.0}$ | $\epsilon_{>1.5}$ |
| 0.50 | 1.00 | 21.00 | 108 | 19.3 | 24.8 | 92.8 | 68.9 | 60.5 | 93.5 |
| 0.50 | 1.00 | 20.00 | 95 | 6.4 | 20.6 | 83.5 | 80.6 | 69.3 | 97.0 |
| 0.50 | 1.00 | 19.00 | 85 | 2.7 | 4.4 | 47.6 | 45.2 | 64.8 | 98.1 |
| 0.50 | 1.00 | 18.75 | 75 | 3.6 | 8.2 | 67.2 | 76.4 | 71.2 | 98.0 |
| 0.50 | 1.00 | 18.50 | 73 | 5.9 | 10.2 | 70.2 | 81.0 | 79.5 | 98.3 |
| 0.50 | 1.00 | 18.25 | 63 | 3.3 | 5.8 | 79.1 | 71.6 | 63.1 | 99.4 |
| 0.50 | 1.00 | 18.00 | 60 | 5.0 | 8.4 | 80.5 | 77.6 | 81.0 | 99.7 |
| 1.00 | 1.50 | 21.00 | 71 | 66.3 | 91.4 | 88.2 | 62.5 | 50.6 | 39.7 |
| 1.00 | 1.50 | 20.00 | 67 | 69.2 | 89.7 | 77.9 | 52.4 | 47.9 | 37.0 |
| 1.00 | 1.50 | 19.00 | 61 | 66.3 | 75.2 | 77.5 | 42.0 | 64.6 | 40.3 |
| 1.00 | 1.50 | 18.75 | 58 | 69.2 | 84.7 | 85.9 | 52.6 | 63.4 | 28.1 |
| 1.00 | 1.50 | 18.50 | 54 | 30.8 | 67.2 | 61.6 | 46.9 | 42.0 | 13.1 |
| 1.00 | 1.50 | 18.25 | 40 | 21.9 | 69.8 | 68.5 | 65.5 | 57.4 | 29.9 |
| 1.00 | 1.50 | 18.00 | 36 | 24.9 | 76.0 | 58.2 | 69.5 | 40.7 | 52.1 |
| 1.50 | 2.00 | 21.00 | 28 | 3.7 | 16.0 | 2.9 | 71.6 | 86.1 | 89.2 |
| 1.50 | 2.00 | 20.00 | 26 | 4.2 | 12.8 | 1.0 | 88.8 | 86.7 | 96.7 |
| 1.50 | 2.00 | 19.00 | 21 | 2.7 | 12.3 | 1.3 | 86.4 | 87.1 | 98.3 |
| 1.50 | 2.00 | 18.75 | 17 | 9.1 | 16.7 | 3.6 | 79.2 | 66.1 | 98.4 |
| 1.50 | 2.00 | 18.50 | 16 | 8.0 | 16.8 | 3.7 | 84.4 | 76.0 | 98.2 |
| 1.50 | 2.00 | 18.25 | 9 | 3.0 | 1.6 | 0.3 | 19.8 | 37.2 | 98.5 |
| 1.50 | 2.00 | 18.00 | 7 | 6.0 | 1.0 | 0.3 | 59.4 | 79.3 | 100.0 |

sources show highly significant correlations with X-ray emission in at least one energy band.

The comparison between results obtained from source and control fields becomes more pronounced in Fig.4. The striking difference between them is immediately apparent, and, unexpected at first sight, the control fields exhibit significant *anti*correlation with X-ray emission. Before we turn to a likely explanation of these anticorrelations, we employ the Kolmogorov-Smirnov test to see whether the distributions of individual error levels from source fields, from control fields, and the theoretically expected distribution of error levels, can significantly be discerned.

## 5 Kolmogorov-Smirnov tests

Given two (normalized, cumulative) distributions $P_{1,2}(x)$ of the same, single independent variable $x$, the Kolmogorov-Smirnov (hereafter KS) test yields the statistical error level $\epsilon_{\rm KS}$ for these distributions to be different (cf. Press et al. 1992). We have three distributions to compare: the distribution of individual correlation coefficients of the source fields, the corresponding distribution for the control fields, and the theoretically expected distribution of correlation coefficients [e.g., Eq.(5) of BS3] for randomly distributed count ranks. The KS test uses the maximum of the absolute value of the difference between the two distributions to be compared. This means that it is efficient in detecting a horizontal shift between the two distributions, especially if they are steep, but less efficient in detecting a deformation of one distribution with respect to the other which leaves the median of the distributions approximately unchanged. We therefore expect that the KS test will be less sensitive to detect significant differences between source and control samples. $\epsilon_{\rm KS}$ is the statistical significance for two distributions to be equal; conversely, it can be considered the error level of the hypothesis that the two distributions differ. Thus, a small value of $\epsilon_{\rm KS}$ indicates a significant difference between the two distributions.

**Table 2.** Correlation results obtained from source and control fields (data displayed in Fig.4). The symbols are defined in the caption of Tab.1

| $z_{\min}$ | $m_{\max}$ | $N$ | source fields | | | control fields | | |
|---|---|---|---|---|---|---|---|---|
| | | | $\epsilon_{\geq 0.75}$ | $\epsilon_{\geq 1.0}$ | $\epsilon_{\geq 1.5}$ | $\epsilon_{\geq 0.75}$ | $\epsilon_{\geq 1.0}$ | $\epsilon_{\geq 1.5}$ |
| | | | (in %) | | | (in %) | | |
| 0.75 | 21.00 | 179 | 13.3 | 13.6 | 38.7 | 95.4 | 91.0 | 88.0 |
| 0.75 | 20.00 | 166 | 6.0 | 5.6 | 19.7 | 95.8 | 93.5 | 89.2 |
| 0.75 | 19.00 | 144 | 4.0 | 4.7 | 10.2 | 86.5 | 92.3 | 86.0 |
| 0.75 | 18.75 | 126 | 7.2 | 8.9 | 31.8 | 85.3 | 92.5 | 95.8 |
| 0.75 | 18.50 | 120 | 3.0 | 5.9 | 16.8 | 84.6 | 83.9 | 95.8 |
| 0.75 | 18.25 | 87 | 0.0 | 1.6 | 17.4 | 70.9 | 65.4 | 91.6 |
| 0.75 | 18.00 | 80 | 0.2 | 1.5 | 9.7 | 88.4 | 77.1 | 96.6 |
| 1.25 | 21.00 | 97 | 19.2 | 56.4 | 20.1 | 94.3 | 96.2 | 83.2 |
| 1.25 | 20.00 | 93 | 29.5 | 40.4 | 26.8 | 92.2 | 92.8 | 81.8 |
| 1.25 | 19.00 | 80 | 12.7 | 25.3 | 9.9 | 85.3 | 90.4 | 88.1 |
| 1.25 | 18.75 | 68 | 12.3 | 38.4 | 27.9 | 73.1 | 84.7 | 87.0 |
| 1.25 | 18.50 | 64 | 7.7 | 32.1 | 11.1 | 63.3 | 78.1 | 88.1 |
| 1.25 | 18.25 | 42 | 4.8 | 7.8 | 12.8 | 67.0 | 81.3 | 87.3 |
| 1.25 | 18.00 | 37 | 10.8 | 13.1 | 15.3 | 94.6 | 90.8 | 98.9 |
| 1.75 | 21.00 | 46 | 10.7 | 15.0 | 28.3 | 98.9 | 98.9 | 84.2 |
| 1.75 | 20.00 | 44 | 4.2 | 8.2 | 33.0 | 98.5 | 97.7 | 89.2 |
| 1.75 | 19.00 | 37 | 2.2 | 4.5 | 13.7 | 98.1 | 93.1 | 83.5 |
| 1.75 | 18.75 | 29 | 0.4 | 5.9 | 23.9 | 94.9 | 91.8 | 92.7 |
| 1.75 | 18.50 | 27 | 0.5 | 5.6 | 20.3 | 92.7 | 89.3 | 88.7 |
| 1.75 | 18.25 | 16 | 0.2 | 0.5 | 6.9 | 77.2 | 85.6 | 94.3 |
| 1.75 | 18.00 | 15 | 0.5 | 1.1 | 7.1 | 86.3 | 90.3 | 85.5 |

We have three different distributions to compare: the distribution of correlation coefficients of individual source- and control fields, $P_{\rm s}(r_{\rm corr})$ and $P_{\rm c}(r_{\rm corr})$, respectively, and the Student-t distribution $P_{\rm t}(r_{\rm corr})$ theoretically expected for randomly distributed photons in the fields. These three distributions are plotted in panel (a) of Fig.5 for $E_\gamma \geq 0.75$ keV, $m_{\max} = 19$, and $z_{\min} = 0.5$.

In addition, Fig.5 displays $\epsilon_{\rm KS}$ as a function of the limiting redshift $z_{\min}$, for photon energies $E_\gamma \geq 0/75$ keV and $m_{\max} \in \{18, 19, 20\}$ (solid, dotted, and dashed lines, respectively). Panel (b) shows $\epsilon_{\rm KS}$ for the comparison of $P_{\rm s}$ with $P_{\rm c}$, panels (c) and (d) for the comparison between $P_{\rm s,c}$ and $P_{\rm t}$. There are redshift ranges exhibiting significant differences between the two distributions, e.g., for $0.5 \lesssim z_{\min} \lesssim 0.8$ and, in particular, for $1.5 \lesssim z_{\min} \lesssim 2.0$. This reflects the aforementioned observation that there is an intermediate redshift interval where 1-Jansky sources show no significant correlation with X-ray emission, and that the most significant correlations occur for high-redshift, optically bright 1-Jansky sources. For $m_{\max} = 18$ (solid curves), there are probably too few data points for the KS test to work reliably.

The anticorrelations found in part of the control-field subsamples also show up in some curves in panel (d) of Fig.5, where the deviation between the distributions comes from the large fraction of small correlation coefficients, in contrast to panel (c) of Fig.5,

**Table 3.** Correlation results obtained from bootstrapping source and control fields; for the definition of the symbols, see the caption of Tab.1

| $z_{\min}$ | $z_{\max}$ | $m_{\max}$ | $N$ | source fields | | | control fields | | |
|---|---|---|---|---|---|---|---|---|---|
| | | | | $\epsilon_{\geq 0.75}$ | $\epsilon_{\geq 1.0}$ | $\epsilon_{\geq 1.5}$ | $\epsilon_{\geq 0.75}$ | $\epsilon_{\geq 1.0}$ | $\epsilon_{\geq 1.5}$ |
| | | | | (in %) | | | (in %) | | |
| 0.50 | 1.00 | 21.00 | 108 | $15.9^{+7.3}_{-4.4}$ | $20.9^{+6.7}_{-4.3}$ | $89.5^{+1.9}_{-1.5}$ | $62.6^{+7.2}_{-4.3}$ | $58.2^{+7.8}_{-5.9}$ | $94.4^{+1.1}_{-2.7}$ |
| 0.50 | 1.00 | 20.00 | 95 | $13.0^{+2.7}_{-2.7}$ | $22.2^{+4.1}_{-4.6}$ | $81.5^{+3.0}_{-1.5}$ | $68.5^{+3.9}_{-21.1}$ | $58.4^{+4.3}_{-2.7}$ | $96.6^{+0.4}_{-0.9}$ |
| 0.50 | 1.00 | 19.00 | 85 | $2.3^{+1.2}_{-0.6}$ | $7.4^{+1.0}_{-1.6}$ | $58.0^{+2.4}_{-7.7}$ | $35.2^{+7.6}_{-6.5}$ | $48.8^{+5.1}_{-8.0}$ | $97.5^{+0.2}_{-0.9}$ |
| 0.50 | 1.00 | 18.75 | 75 | $4.8^{+0.6}_{-1.3}$ | $11.9^{+1.3}_{-3.1}$ | $62.6^{+2.8}_{-6.5}$ | $60.7^{+2.7}_{-4.0}$ | $51.3^{+5.2}_{-4.3}$ | $98.1^{+0.7}_{-0.2}$ |
| 0.50 | 1.00 | 18.50 | 73 | $6.2^{+1.2}_{-1.2}$ | $13.8^{+2.3}_{-1.5}$ | $65.5^{+4.1}_{-2.6}$ | $73.4^{+2.4}_{-11.4}$ | $63.7^{+4.0}_{-20.1}$ | $98.6^{+0.3}_{-0.2}$ |
| 0.50 | 1.00 | 18.25 | 63 | $4.2^{+1.3}_{-0.6}$ | $8.1^{+2.1}_{-0.5}$ | $78.0^{+6.7}_{-3.3}$ | $81.1^{+2.4}_{-9.2}$ | $63.7^{+8.1}_{-11.8}$ | $98.5^{+0.4}_{-0.4}$ |
| 0.50 | 1.00 | 18.00 | 60 | $6.4^{+0.9}_{-0.9}$ | $6.7^{+1.9}_{-1.0}$ | $77.3^{+0.8}_{-1.4}$ | $89.5^{+1.0}_{-6.2}$ | $74.5^{+2.7}_{-16.0}$ | $99.4^{+0.7}_{-3.1}$ |
| 1.00 | 1.50 | 21.00 | 71 | $61.5^{+5.3}_{-7.1}$ | $93.8^{+0.0}_{-5.1}$ | $86.6^{+1.6}_{-2.4}$ | $58.5^{+5.2}_{-12.1}$ | $51.6^{+2.2}_{-6.0}$ | $36.2^{+3.7}_{-5.1}$ |
| 1.00 | 1.50 | 20.00 | 67 | $72.3^{+7.2}_{-6.8}$ | $89.7^{+1.8}_{-1.8}$ | $83.6^{+2.7}_{-2.0}$ | $51.0^{+2.4}_{-3.3}$ | $49.3^{+6.5}_{-2.9}$ | $31.2^{+3.7}_{-10.7}$ |
| 1.00 | 1.50 | 19.00 | 61 | $62.1^{+4.6}_{-2.6}$ | $78.1^{+7.4}_{-3.4}$ | $72.9^{+9.5}_{-1.8}$ | $39.1^{+5.0}_{-3.6}$ | $58.0^{+7.2}_{-3.8}$ | $25.5^{+3.2}_{-1.1}$ |
| 1.00 | 1.50 | 18.75 | 58 | $66.2^{+3.9}_{-5.4}$ | $85.6^{+2.4}_{-3.3}$ | $78.6^{+3.2}_{-2.4}$ | $60.9^{+1.6}_{-2.1}$ | $66.1^{+6.3}_{-4.4}$ | $17.9^{+7.6}_{-1.7}$ |
| 1.00 | 1.50 | 18.50 | 54 | $28.5^{+6.6}_{-5.8}$ | $64.0^{+2.2}_{-4.4}$ | $64.8^{+4.2}_{-3.0}$ | $44.1^{+4.7}_{-3.5}$ | $38.4^{+2.9}_{-0.3}$ | $14.2^{+5.5}_{-1.8}$ |
| 1.00 | 1.50 | 18.25 | 40 | $21.1^{+3.5}_{-1.8}$ | $67.0^{+2.7}_{-8.1}$ | $65.6^{+0.8}_{-4.8}$ | $61.4^{+5.6}_{-2.0}$ | $56.6^{+3.8}_{-6.7}$ | $26.1^{+4.8}_{-2.2}$ |
| 1.00 | 1.50 | 18.00 | 36 | $18.0^{+6.3}_{-3.2}$ | $64.2^{+5.6}_{-1.2}$ | $51.2^{+5.0}_{-7.3}$ | $64.7^{+3.3}_{-4.5}$ | $39.2^{+5.4}_{-4.0}$ | $39.6^{+0.9}_{-7.8}$ |
| 1.50 | 2.00 | 21.00 | 28 | $2.6^{+0.9}_{-0.4}$ | $10.4^{+1.8}_{-1.7}$ | $1.4^{+0.1}_{-0.4}$ | $88.8^{+1.1}_{-5.3}$ | $87.5^{+2.2}_{-4.0}$ | $83.7^{+1.1}_{-10.4}$ |
| 1.50 | 2.00 | 20.00 | 26 | $2.4^{+0.9}_{-0.7}$ | $6.4^{+0.9}_{-1.1}$ | $0.7^{+1.0}_{-0.3}$ | $94.4^{+0.4}_{-0.6}$ | $92.1^{+1.3}_{-1.0}$ | $92.3^{+1.5}_{-0.6}$ |
| 1.50 | 2.00 | 19.00 | 21 | $2.1^{+0.3}_{-0.0}$ | $7.9^{+0.8}_{-1.1}$ | $1.8^{+0.4}_{-0.2}$ | $89.0^{+1.2}_{-1.2}$ | $91.8^{+1.4}_{-2.4}$ | $95.9^{+1.0}_{-2.1}$ |
| 1.50 | 2.00 | 18.75 | 17 | $4.3^{+0.3}_{-0.3}$ | $8.0^{+0.5}_{-0.0}$ | $3.6^{+1.0}_{-0.5}$ | $86.5^{+2.8}_{-0.6}$ | $78.2^{+2.6}_{-1.1}$ | $96.4^{+0.7}_{-0.5}$ |
| 1.50 | 2.00 | 18.50 | 16 | $3.9^{+2.7}_{-0.4}$ | $9.1^{+1.6}_{-0.3}$ | $1.9^{+0.9}_{-0.1}$ | $84.5^{+1.8}_{-1.8}$ | $82.7^{+2.8}_{-0.9}$ | $94.9^{+0.9}_{-0.9}$ |
| 1.50 | 2.00 | 18.25 | 9 | $0.4^{+0.0}_{-0.1}$ | $0.5^{+0.1}_{-0.1}$ | $0.2^{+0.0}_{-0.0}$ | $31.5^{+5.6}_{-2.3}$ | $59.2^{+0.7}_{-2.2}$ | $97.0^{+0.6}_{-0.4}$ |
| 1.50 | 2.00 | 18.00 | 7 | $0.5^{+0.4}_{-0.0}$ | $0.1^{+0.0}_{-0.0}$ | $0.3^{+0.1}_{-0.0}$ | $90.6^{+1.1}_{-2.9}$ | $84.0^{+0.0}_{-1.5}$ | $99.3^{+0.0}_{-0.2}$ |

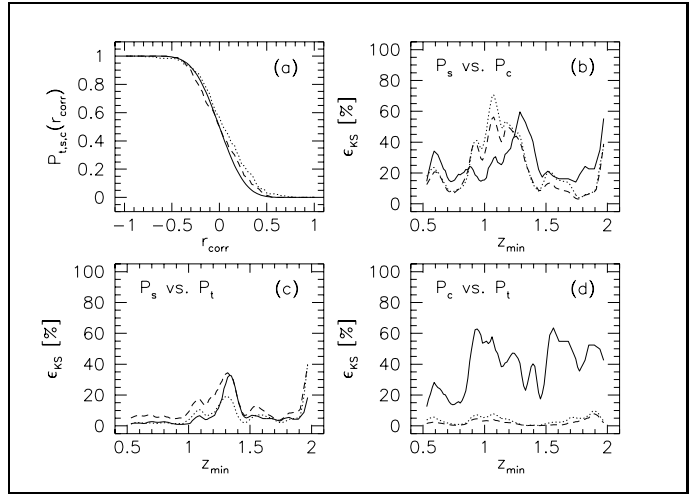

**Fig. 5.** Panel (a): the distributions $P_{\mathrm{t,s,c}}$ (solid, dotted, and dashed lines, respectively) for $E_\gamma \geq 0.75$ keV, $m_{\max} = 19$, and $z_{\min} = 0.5$; panels (b,c,d): results $\epsilon_{\mathrm{KS}}(z_{\min})$ of a KS comparison between pairs of these distributions. The abscissae are limiting source redshifts $z_{\min}$, i.e., from left to right the source subsamples become more distant. The ordinate is the error level for the distributions to be different, i.e., the significance for them to be different increases downward. The three curves per panel are for different optical threshold magnitudes of the source subsamples, namely for $m_{\max} \in \{18, 19, 20\}$ (solid, dotted, and dashed lines, respectively)

where the reason for the significant deviations is the occurrence of very large correlation coefficients in the source sample. Note that the results for $m_{\max} = 18$ are again the least significant because of the small numbers of sources.

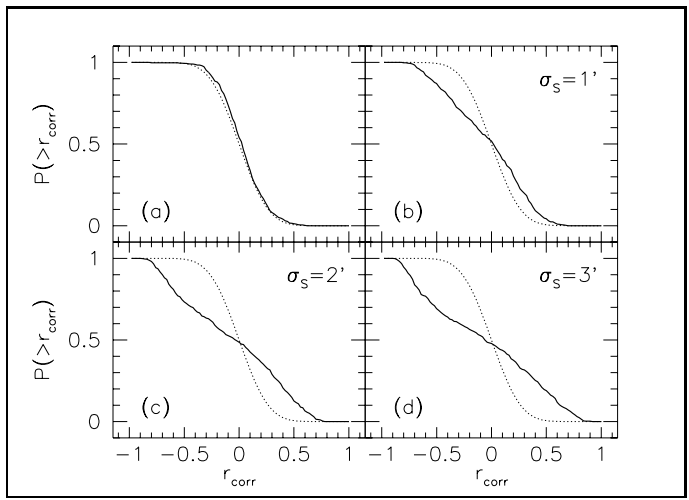

**Fig. 6.** Simulated cumulative distributions of rank-order correlation coefficients. In all four panels, the dotted curve displays the theoretically expected distribution for randomly distributed ranks. The simulations for panel (a) were purely random fields, while in the simulations for the other panels, we have added sources with different widths $\sigma_s$ to one half of the fields ($f = 0.5$)

## 6 Origin of anticorrelations in control fields

At first sight, the highly significant anticorrelations found in many of the control-field subsamples (see, in particular, panel (d) of Fig.5 and Tab.1) are a puzzling result. Looking into a randomly selected direction on the sky, a random portion of the X-ray emission should be seen on average, and it is therefore striking to find the emission anticorrelated with the randomly chosen control-field centers.

The explanation, however, is rather straightforward. The X-ray emission is due to both diffuse and discrete sources, where the image of such a discrete source can have a diameter of even a few arc minutes dependent on its brightness; bright sources appear more extended than dim sources[2]. Consider now a field containing randomly distributed photon events, and, in addition, the photons from randomly placed discrete sources with given extension. For the purposes of rank-order statistics, this field is then covered with 25 ring-shaped, concentric cells. Because these cells are chosen to have equal area, they become thinner outward. When the radius of the discrete source becomes of the order of $\simeq 1'$, its photons will affect ('contaminate') several adjacent cells if the source is close to the outer boundary of the cell pattern, but only one or few cells if it is close to the center. Therefore, once the source appears large enough to cover several cells at the boundary, it produces a coherent 'distortion' of the count ranks of distant cells, but it affects only single cells close to the center. In a sample of such fields, the contamination of single cells close to the center is likely to be 'averaged away', but not the coherent contamination of multiple cells close to the outer boundary. An example of a discrete source close to the outer cell-pattern boundary is seen in panel (j) of Fig.2.

To see whether this qualitative consideration can indeed account for the anticorrelations detected in the control fields, we have performed Monte-Carlo simulations, as follows.

In a sample of 100 circular fields with radius $\sqrt{100 \times 25/\pi}\ ' \simeq 28'.2$, $N_p$ points (representing photons) are distributed following a Poisson distribution with $\langle N_p \rangle = 500$. Additionally, a fraction $f$ of these fields are contaminated with randomly placed 'compact sources' of Gaussian profile with variance taken randomly from the interval $[0.5\sigma_\mathrm{s}, 1.5\sigma_\mathrm{s}]$ to mimic different source diameters, where $\sigma_\mathrm{s}$ is taken as a parameter; the number of 'photons' per 'source' was randomly drawn from the interval $[20, 500]$. This sample is then subjected to rank-order statistics alike the source and control fields. Fig.6 shows the cumulative probability distribution for the correlation coefficients $r_\mathrm{corr}$ obtained from 100 runs with $f = 0.5$ and compares them with the theoretical distribution expected for randomly distributed ranks.

---

[2] Additionally, the point-spread function depends on the photon energy; this dependence is, however, negligible compared to its broadening for off-axis sources. In contrast to pointed observations, the spatial photon distribution of compact sources in the *All-Sky Survey* is dominated by the shape of the off-axis PSF.

The figure shows that, for uncontaminated fields, the simulated distribution closely follows the theoretical curve (upper left panel), but the deviations between the curves increase with increasing average width $\sigma_\mathrm{s}$ of the contaminating sources. Table 4 gives, for 10 different simulations (each containing 100 'source fields') and four values of $\sigma_\mathrm{s}$, the rank-order correlation error levels $\epsilon$ together with the Kolmogorov-Smirnov error levels $\epsilon_\mathrm{KS}$ from comparing with $P_\mathrm{t}$. $\sigma_\mathrm{s} = 0$ abbreviates simulations without contamination by discrete sources.

**Table 4.** Results from ten independent Monte-Carlo simulations of $N_\mathrm{field} = 100$ random fields, to half of which ($f = 0.5$) discrete sources were added. $\sigma_\mathrm{s}$ is the average source diameter in arc minutes, $\epsilon$ is the rank-order correlation error level, and $\epsilon_\mathrm{KS}$ is the KS error level, both in %. $\sigma_\mathrm{s} = 0$ means no contamination by sources

| simulation # | $\sigma_\mathrm{s}$ (arcmin) | | | | | | | |
| --- | --- | --- | --- | --- | --- | --- | --- | --- |
| | 0 | | 1 | | 2 | | 3 | |
| | $\epsilon$ | $\epsilon_\mathrm{KS}$ | $\epsilon$ | $\epsilon_\mathrm{KS}$ | $\epsilon$ | $\epsilon_\mathrm{KS}$ | $\epsilon$ | $\epsilon_\mathrm{KS}$ |
| 1 | 22.47 | 48.11 | 69.45 | 2.03 | 99.90 | 0.00 | 71.99 | 0.03 |
| 2 | 9.07 | 25.24 | 99.78 | 0.03 | 99.67 | 0.00 | 96.89 | 0.00 |
| 3 | 78.09 | 42.80 | 94.24 | 3.03 | 62.75 | 0.33 | 93.27 | 0.00 |
| 4 | 46.80 | 89.40 | 70.86 | 1.23 | 5.02 | 0.00 | 99.99 | 0.00 |
| 5 | 4.94 | 37.48 | 99.92 | 0.01 | 97.09 | 0.00 | 97.25 | 0.00 |
| 6 | 23.48 | 84.36 | 12.95 | 0.12 | 66.83 | 0.04 | 99.99 | 0.00 |
| 7 | 33.44 | 59.00 | 35.19 | 8.56 | 92.64 | 0.24 | 99.66 | 0.00 |
| 8 | 47.09 | 43.97 | 92.85 | 0.22 | 99.98 | 0.00 | 100.00 | 0.00 |
| 9 | 80.54 | 49.63 | 83.30 | 0.29 | 86.06 | 0.02 | 100.00 | 0.00 |
| 10 | 22.80 | 72.18 | 73.94 | 0.06 | 86.15 | 0.07 | 97.33 | 0.00 |

The table quantifies the results displayed in Fig.6 and shows that, with increasing average source width, anticorrelations become more frequent and more significant ($\epsilon \to 100\%$), and the deviation of the distribution of correlation coefficients from the theoretical distribution becomes highly significant ($\epsilon_\mathrm{KS} \to 0\%$), just as described above.

Moreover, we show in Fig.7 the change in the KS error level $\epsilon_\mathrm{KS}$ with decreasing number $N_\mathrm{field}$ of simulated fields. The figure confirms that, with increasing source size, the correlation-coefficient distribution deviates more and more significantly from the theoretically expected distribution, and it also shows that the deviations become less significant with decreasing subsample size. The two panels in Fig.7 were obtained with $f = 0.5$ (left panel) and $f = 0.25$ (right panel); i.e., in the right panel, the number of contaminated fields is smaller.

These simulations confirm the suspicion that the detected anticorrelations in the control fields originate from the contamination of the fields by the emission from discrete sources which appear extended because of the PSPC point-spread function. This also means that the correlations found in the source-field subsamples occur *in spite* of this effect, and therefore strengthens the significance of these correlations.

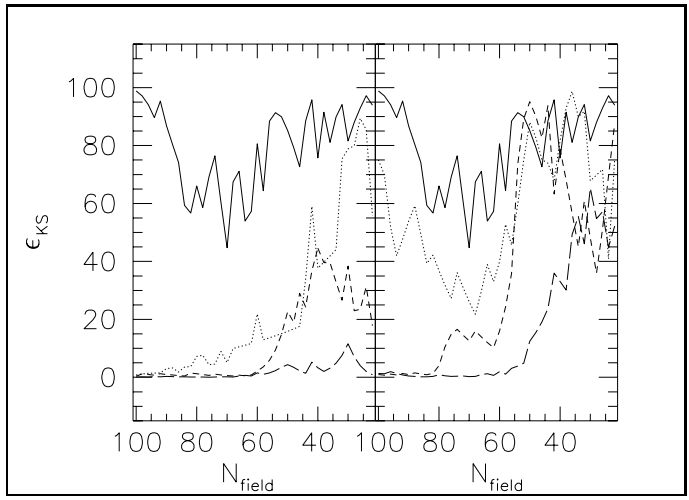

**Fig. 7.** The figure shows Kolmogorov-Smirnov error levels obtained from simulated fields. The fields contained either only randomly distributed points (solid lines) or were additionally contaminated with discrete sources of width $\sigma_s \in \{1,2,3\}'$ (dotted, short-dashed, and dashed lines, respectively) as described in the text. In the left panel, half of the simulated fields contained such sources, in the right panel, one quarter.

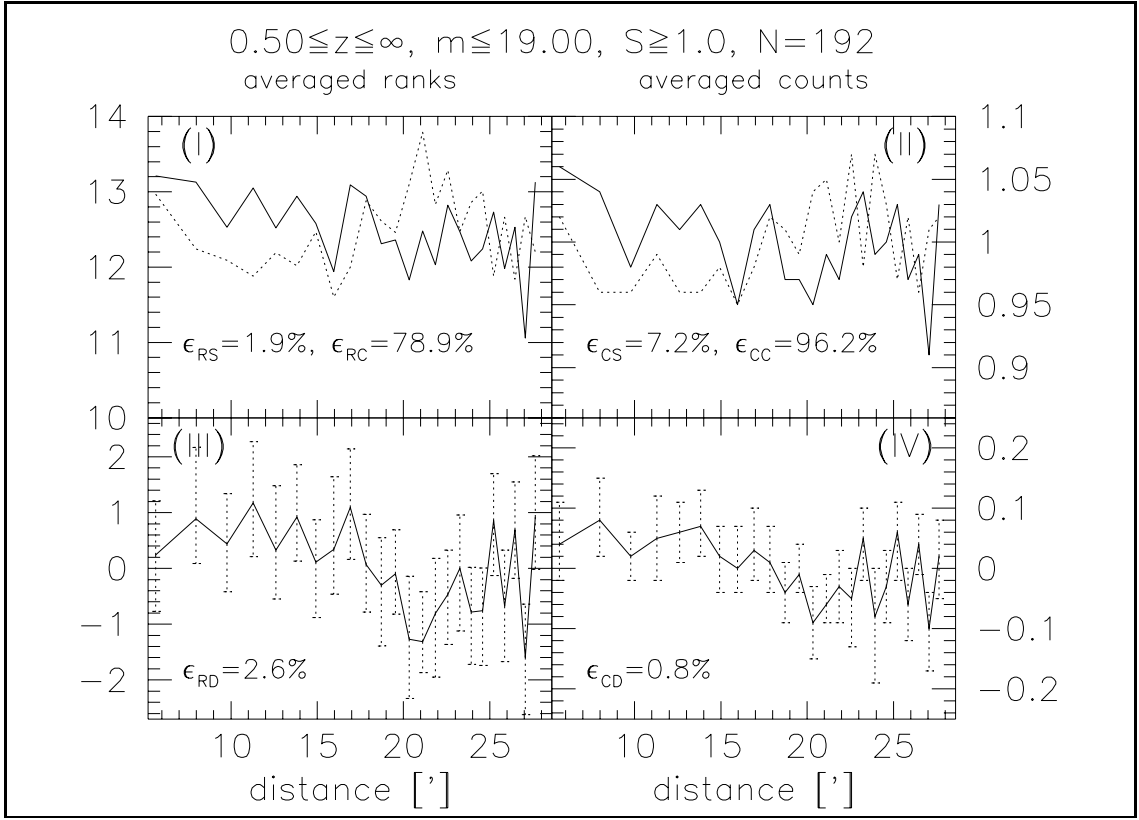

**Fig. 8a.** For the subsample defined by $z \geq 0.5$, $m \leq 19$ and radio flux $S \geq 1$ Jy, we have plotted in panel (I) the mean ranks $\langle \mathcal{R}_S(j) \rangle$ (solid curve) and $\langle \mathcal{R}_C(j) \rangle$ (dashed curve) of the source and control fields, and their difference $\Delta \langle \mathcal{R}(j) \rangle$ in panel (III). Panel (II) show the normalized average counts $\bar{C}_S(j)$ (solid curve) and $\bar{C}_C(j)$ (dashed curve) of the source and control fields, respectively, and their difference $\Delta \bar{C}(j)$ in panel (IV). In each panel, the corresponding error levels from rank-order statistics are indicated. The error bars in panels (III) and (IV) are bootstrap error bars obtained as described in Sect.4, i.e., they indicate the range containing 80% of all bootstrap-resampled data sets

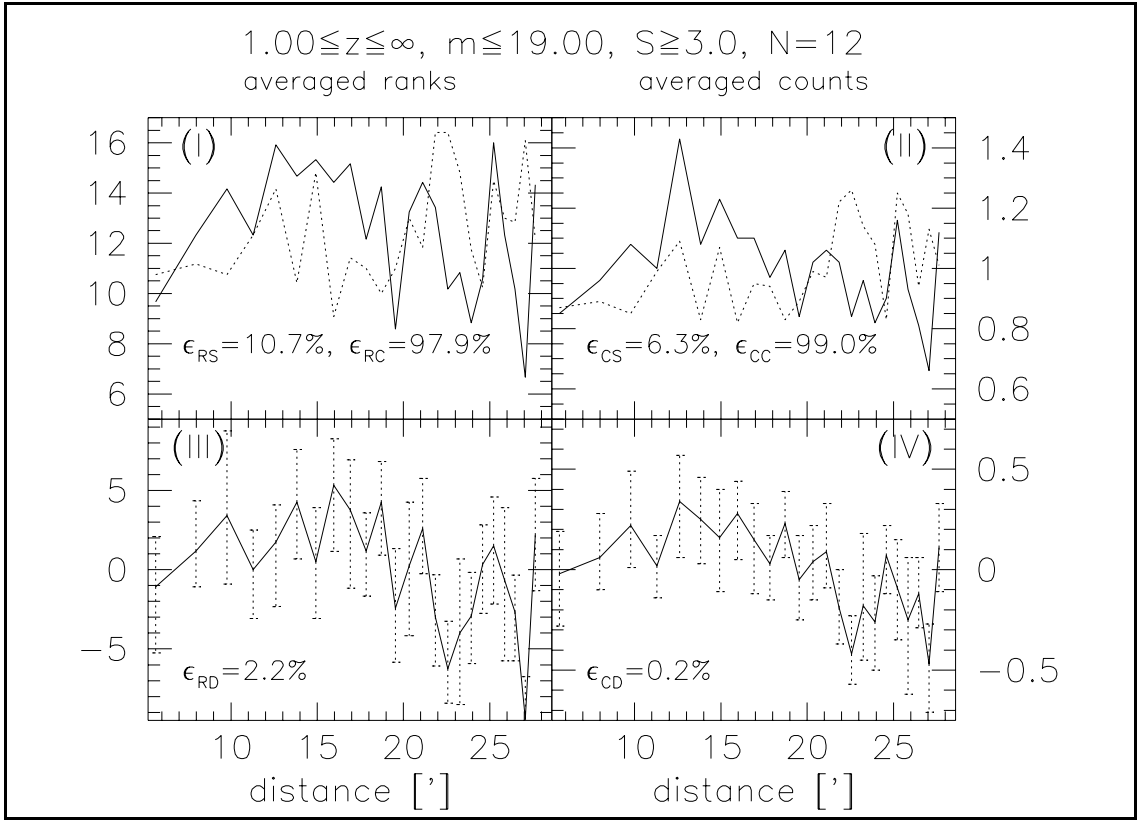

**Fig. 8b.** Same as Fig.8a, for a subsample with $z \geq 1$, $m \leq 19$ and radio flux $S \geq 3\,\mathrm{Jy}$

## 7 Further statistical tests

Following the preceding remark, we want to further investigate the statistical differences between the photon counts in the source fields and their control fields. The statistical analysis presented in Sect.4 was made by ranking the cell pattern of each source and each control field (for a chosen subsample), then averaging over all sources and control fields of this subsample. The resulting average rank schemes $\langle \mathcal{R}(j) \rangle$ for the source and control fields were then again ranked, and a correlation coefficient and the corresponding error level were determined. Henceforth, we denote the error levels for the source and control fields by $\epsilon_{\mathrm{RS}}$ and $\epsilon_{\mathrm{RC}}$, respectively. Examples for the mean ranks of several subsamples are found in panels (I) of Figs.8a,b,c; the solid and dotted curves correspond to the source and control samples, respectively. Panels (III) of these figures show the difference

$$\Delta \langle \mathcal{R}(j) \rangle = \langle \mathcal{R}_{\mathrm{S}}(j) \rangle - \langle \mathcal{R}_{\mathrm{C}}(j) \rangle \tag{1}$$

of the ranks of the source fields and the control fields. We have also ranked this difference rank scheme $\Delta \langle \mathcal{R}(i) \rangle$ and obtained a correlation coefficient and, correspondingly, an error level, which we denote by $\epsilon_{\mathrm{RD}}$.

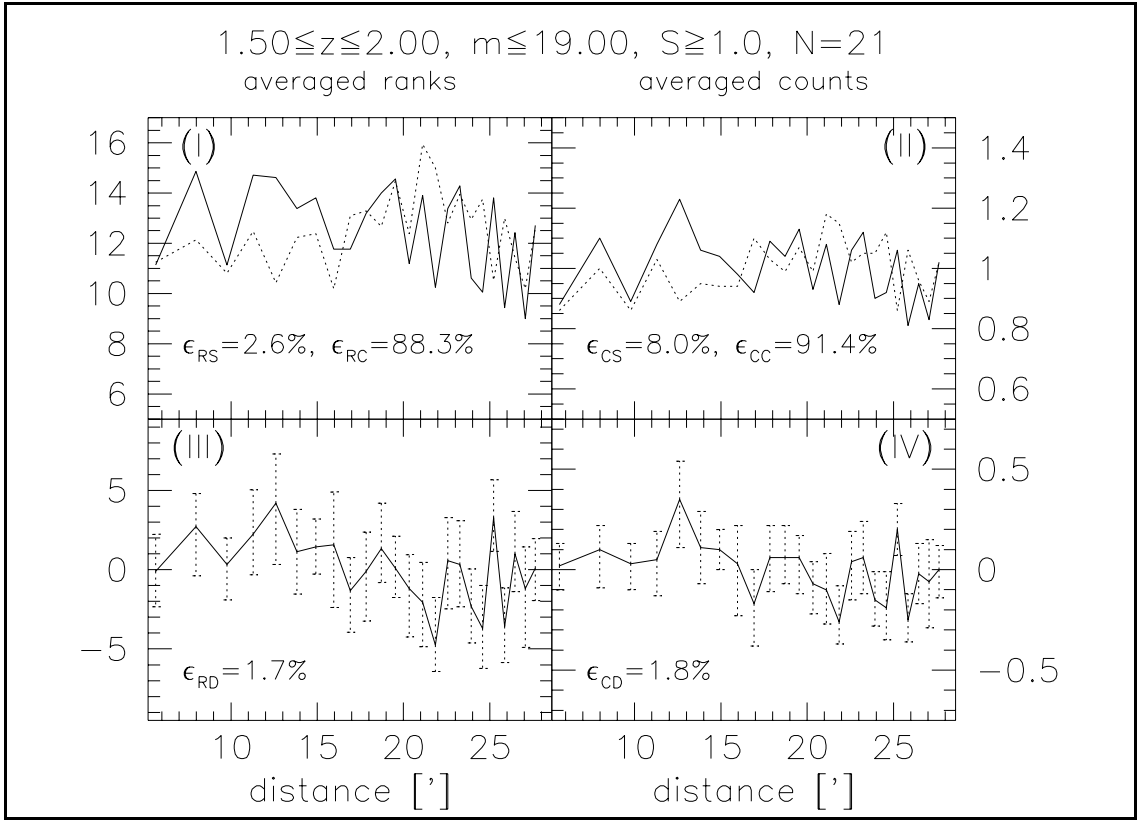

**Fig. 8c.** Same as Fig.8a, for a subsample with $1.5 \leq z \leq 2$, $m \leq 19$ and radio flux $S \geq 1\,\mathrm{Jy}$

In addition, we have calculated the mean photon counts of all source and control fields of a chosen subsample, weighted by the total photon counts in each source and control field in the rings 2 to 25. More precisely, if $C(i,j)$ denotes the photon counts in the $j$-th ring of the $i$-th source (or control) field of a chosen subsample, we have calculated

$$\bar{C}(j) = \sum_i \frac{C(i,j)}{\frac{1}{24}\sum_{j=2}^{25} C(i,j)} \; ; \qquad (2)$$

these normalized average counts are plotted in panels (II) of Figs.8a,b,c for several subsamples, for both the source fields (solid curves) and the control fields (dashed curves). We have ranked the $\bar{C}(j)$ and obtained the corresponding error levels $\epsilon_{\mathrm{CS}}$ and $\epsilon_{\mathrm{CC}}$ for the source and control fields, respectively. In addition, we have ranked the difference

$$\Delta\bar{C}(j) = \bar{C}_{\mathrm{S}}(j) - \bar{C}_{\mathrm{C}}(j) \; , \qquad (3)$$

plotted in panels (IV) of these figures, and obtained the error level $\epsilon_{\mathrm{CD}}$. In Table 5, we list the error levels introduced above for several subsamples.

Several trends are worth mentioning: as expected, the error levels for the differences, both for $\Delta\langle\mathcal{R}(j)\rangle$ and for $\Delta\bar{C}(j)$ are smaller than $\epsilon_{\mathrm{RS}}$ and $\epsilon_{\mathrm{CS}}$, since the anticorrelation of the control fields, which we have attributed to the existence of (supposedly unrelated) discrete sources in the fields, must also be present in the source fields. By taking the difference between source and control fields, this effect should be largely eliminated from the source fields, and the resulting schemes should be cleaner representations of the truly correlated counts. Second, the error levels obtained from the mean ranks and the normalized average counts are comparable, i.e., they do not differ significantly. The figures give a good illustration of our method, as well as its sensitivity. The correlation of the X-ray photons with the radio QSOs can be seen 'by eye' in these plots, best of course in the plots for the difference of the ranks and the difference of the normalized average counts.

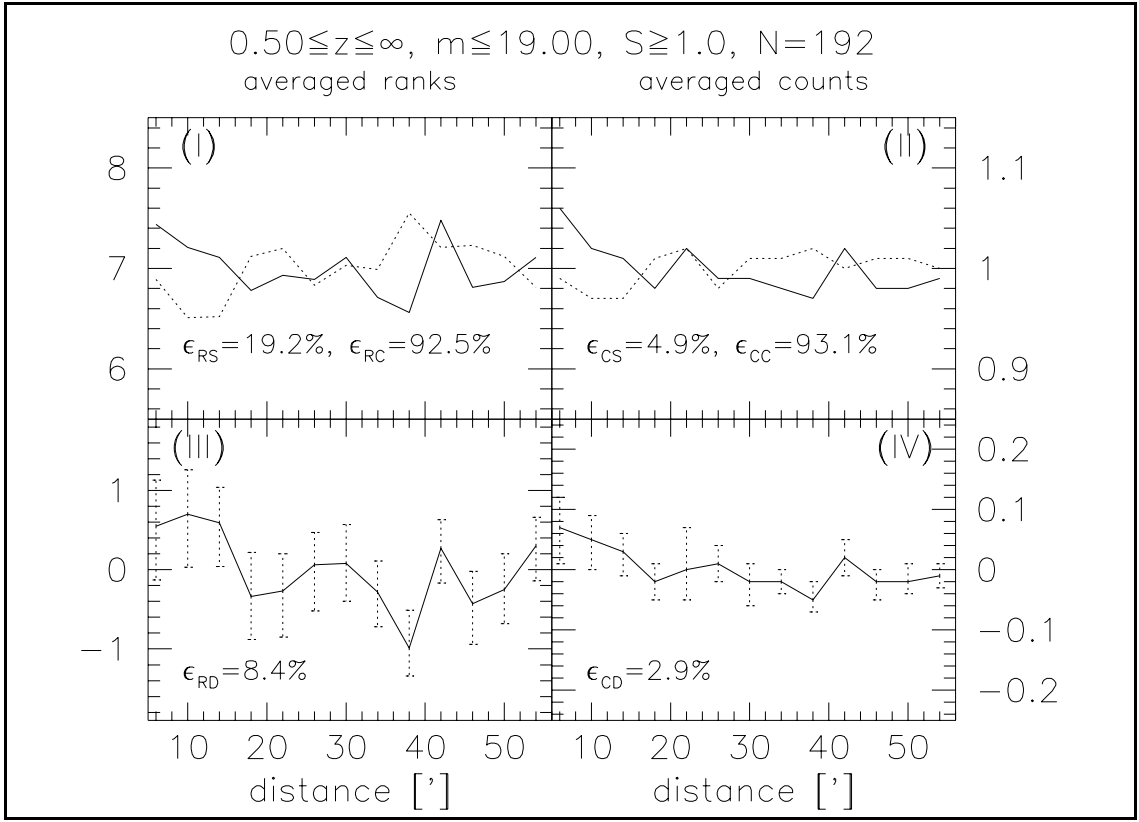

**Fig. 8d.** Similar to Fig.8a, but for twice the angular scale ($5'.6\ldots 57'.6$). The curves, in particular those in the lower panels, are approximately flat for distances $\gtrsim 20'$, and rise for distances $\lesssim 15'$. The distance where the lower curves drop to zero, $\simeq 15'$, roughly corresponds to the Abell radius of galaxy clusters at intermediate redshifts

The correlation results should become less significant when the correlation scale is enlarged. To test this, we have counted X-ray photons in source and control fields in 13 cells of equal *width* (we have chosen $4'$) covering a ring between $5'.6$ and $57'.6$ centered on the field centers, and weighted the counts by the cell area. To these counts, we applied the same type of analysis as described above; the results for one subsample (the same as in Fig.8a) are displayed in Fig.8d. The curves are approximately flat for large distances ($\gtrsim 20'$) from the center, while they rise for distances smaller than $\simeq 15'$. For further discussion, see Sect.9.1.

In order to exclude the possibility that the observed correlations are due to either the wings of the point spread function or to a halo around the QSOs caused by scattering of dust in the interstellar medium, we have repeated the correlation calculations by subtracting a fraction $f$ of the photon counts in the central cell ($j = 1$) from the other 24 rings, according to an angular dependence $\propto \theta^{-\alpha}$. We have tested the cases $f = 1\%$ and $f = 3\%$, and $\alpha = 1$ and $\alpha = 2$; these should provide overestimates of the possible flux from the QSOs spread into the outer rings (see also Fig.1). Whereas the correlation coefficients become slightly smaller if we allow for this spreading, in no case does a strong correlation become significantly weakened. We therefore conclude that the observed correlation is not caused by the wings of the point spread function or a scattering halo around the QSOs.

**Table 5.** Correlation results for the subsamples of Tab.2, obtained from the methods described in Sect.7 applied to the energy regime $E_\gamma \geq 0.75$ keV. $\epsilon_{R(S,C,D)}$ are error levels found from **r**anking the **s**ource and **c**ontrol fields and their **d**ifferences, respectively, while $\epsilon_{C(S,C,D)}$ were derived from normalized photon counts

| $z_{\min}$ | $m_{\max}$ | $S_{\min}$ | $N$ | $\epsilon_{RS}$ | $\epsilon_{RC}$ | $\epsilon_{CS}$ | $\epsilon_{CC}$ | $\epsilon_{RD}$ | $\epsilon_{CD}$ |
|---|---|---|---|---|---|---|---|---|---|
| | | | | | | in (%) | | | |
| 0.75 | 21.00 | 1.0 | 179 | 13.3 | 93.5 | 17.8 | 97.8 | 6.2 | 1.4 |
| 0.75 | 20.00 | 1.0 | 166 | 6.9 | 91.2 | 15.0 | 99.1 | 5.2 | 1.2 |
| 0.75 | 19.00 | 1.0 | 144 | 3.7 | 88.6 | 10.3 | 94.0 | 4.4 | 4.0 |
| 0.75 | 18.75 | 1.0 | 126 | 7.0 | 86.4 | 8.6 | 92.9 | 4.9 | 8.2 |
| 0.75 | 18.50 | 1.0 | 120 | 2.8 | 84.6 | 7.3 | 90.8 | 2.5 | 7.5 |
| 0.75 | 18.25 | 1.0 | 87 | 0.0 | 70.2 | 1.2 | 63.1 | 1.4 | 2.2 |
| 0.75 | 18.00 | 1.0 | 80 | 0.2 | 88.4 | 1.3 | 75.8 | 0.4 | 1.0 |
| 1.25 | 21.00 | 1.0 | 97 | 19.2 | 94.1 | 57.2 | 99.2 | 5.3 | 5.0 |
| 1.25 | 20.00 | 1.0 | 93 | 29.5 | 91.6 | 45.7 | 96.4 | 9.7 | 7.3 |
| 1.25 | 19.00 | 1.0 | 80 | 12.7 | 85.7 | 36.7 | 92.7 | 5.2 | 12.2 |
| 1.25 | 18.75 | 1.0 | 68 | 14.8 | 73.1 | 39.8 | 76.1 | 7.3 | 19.6 |
| 1.25 | 18.50 | 1.0 | 64 | 8.6 | 63.7 | 29.1 | 71.2 | 6.7 | 20.0 |
| 1.25 | 18.25 | 1.0 | 42 | 4.8 | 67.6 | 22.6 | 65.1 | 3.4 | 28.4 |
| 1.25 | 18.00 | 1.0 | 37 | 12.1 | 95.0 | 24.3 | 85.1 | 4.7 | 32.2 |
| 1.75 | 21.00 | 1.0 | 46 | 10.7 | 99.0 | 30.2 | 99.9 | 0.5 | 0.1 |
| 1.75 | 20.00 | 1.0 | 44 | 3.9 | 99.6 | 16.3 | 99.4 | 0.3 | 0.3 |
| 1.75 | 19.00 | 1.0 | 37 | 2.2 | 98.6 | 7.5 | 99.0 | 0.2 | 1.3 |
| 1.75 | 18.75 | 1.0 | 29 | 0.4 | 94.6 | 2.0 | 88.4 | 0.2 | 1.9 |
| 1.75 | 18.50 | 1.0 | 27 | 0.6 | 92.7 | 2.6 | 86.6 | 0.1 | 0.9 |
| 1.75 | 18.25 | 1.0 | 16 | 0.2 | 80.6 | 2.1 | 49.2 | 0.3 | 3.4 |
| 1.75 | 18.00 | 1.0 | 15 | 0.6 | 87.0 | 3.2 | 61.6 | 0.6 | 5.2 |

Another comparison between source and control fields is illustrative: in Fig.9 we have plotted a histogram of the ratio $(N_{\rm pos} - N_{\rm neg})/(N_{\rm pos} + N_{\rm neg})$, where $N_{\rm pos}$ and $N_{\rm neg}$ are the number of fields in which the correlation coefficient in the source field is larger or smaller, respectively, than in the corresponding control field, for different redshift bins. The trend seen in the preceding tables is also visible here: the correlation coefficient in the source fields is larger than that in the corresponding control fields in the majority of cases, if the redshift is either smaller than about 1 or larger than about 1.5, whereas no clear trend is visible in the intermediate redshift regime.

In accordance with the results in Sect.4, the correlation analyses presented here do not yield a significant correlation of QSOs in the redshift interval $1 \leq z \leq 1.5$ with X-ray photons. If the correlations that we observe for other redshifts are due to a magnification bias caused by lensing, we expect the correlation coefficients to increase if we increase the radio flux threshold of the QSO sample, since the radio source counts are steeper for the highest fluxes. Thus we repeated the statistical analysis by considering only QSOs with radio flux larger than 2 or 3 Jy; the results for the redshift interval $1 \leq z \leq 1.5$ are displayed in Tab.6. As one can see, increasing the radio flux threshold of the sample, the correlations, which can not be detected at a significant level for a threshold of 1 Jy, become visible for a larger radio flux threshold. This trend is present in all subsamples, unless the number of sources in the subsample becomes smaller than about 5. Therefore, we conclude that the trend of the correlation coefficient, or the corresponding error level, is in agreement with expectations based on a gravitational lens interpretation of our results. A discussion will be presented in Sect.9.

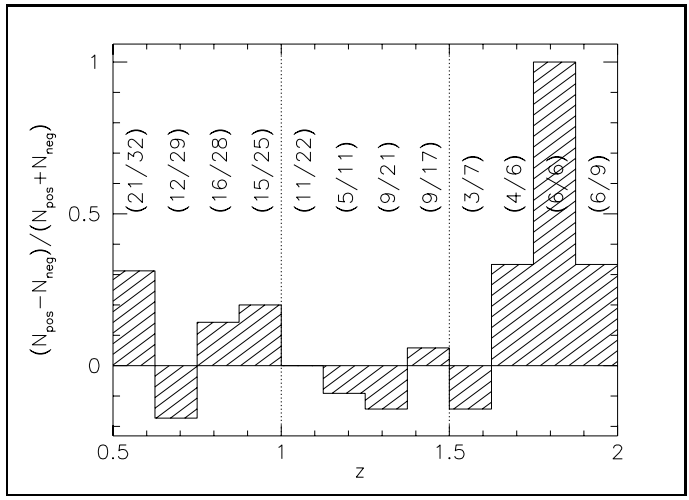

**Fig. 9.** A histogram of the ratio $(N_{\text{pos}} - N_{\text{neg}})/(N_{\text{pos}} + N_{\text{neg}})$, where $N_{\text{pos}}$ and $N_{\text{neg}}$ are the number of fields in which the correlation coefficient in the source field is larger or smaller, respectively, than in the corresponding control field, for different redshift bins. The pair of numbers in each bin denotes $(N_{\text{pos}}/N_{\text{pos}} + N_{\text{neg}})$

**Table 6.** Correlation results for the subsamples of Tab.1, using the methods of Sect.7. For the meaning of the symbols, see the caption of Tab.5

| $z_{\min}$ | $z_{\max}$ | $m_{\max}$ | $S_{\min}$ | $N$ | $\epsilon_{\text{RS}}$ | $\epsilon_{\text{RC}}$ | $\epsilon_{\text{CS}}$ | $\epsilon_{\text{CC}}$ | $\epsilon_{\text{RD}}$ | $\epsilon_{\text{CD}}$ |
|---|---|---|---|---|---|---|---|---|---|---|
| | | | | | \multicolumn{6}{c}{in (%)} | | | | | |
| 1.00 | 1.50 | 21.00 | 1.0 | 71 | 66.3 | 62.5 | 58.3 | 54.5 | 50.2 | 52.1 |
| 1.00 | 1.50 | 20.00 | 1.0 | 67 | 69.2 | 55.3 | 65.7 | 37.8 | 43.3 | 66.7 |
| 1.00 | 1.50 | 19.00 | 1.0 | 61 | 63.7 | 39.8 | 37.8 | 45.0 | 41.2 | 51.9 |
| 1.00 | 1.50 | 18.75 | 1.0 | 58 | 69.5 | 52.6 | 46.0 | 49.0 | 31.8 | 51.1 |
| 1.00 | 1.50 | 18.50 | 1.0 | 54 | 29.5 | 47.3 | 30.4 | 40.8 | 35.5 | 35.5 |
| 1.00 | 1.50 | 18.25 | 1.0 | 40 | 21.9 | 61.9 | 20.1 | 48.4 | 15.9 | 22.3 |
| 1.00 | 1.50 | 18.00 | 1.0 | 36 | 21.7 | 70.0 | 20.0 | 57.5 | 28.8 | 15.1 |
| 1.00 | 1.50 | 21.00 | 2.0 | 18 | 7.7 | 83.6 | 37.5 | 84.1 | 7.5 | 18.9 |
| 1.00 | 1.50 | 20.00 | 2.0 | 16 | 15.6 | 62.8 | 50.2 | 62.2 | 14.5 | 21.2 |
| 1.00 | 1.50 | 19.00 | 2.0 | 16 | 13.9 | 59.1 | 50.2 | 62.2 | 14.6 | 21.2 |
| 1.00 | 1.50 | 18.75 | 2.0 | 16 | 13.9 | 64.9 | 50.2 | 62.2 | 14.2 | 21.2 |
| 1.00 | 1.50 | 18.50 | 2.0 | 15 | 13.0 | 63.3 | 50.0 | 70.3 | 14.5 | 21.1 |
| 1.00 | 1.50 | 18.25 | 2.0 | 12 | 15.2 | 67.9 | 46.6 | 64.2 | 22.4 | 22.6 |
| 1.00 | 1.50 | 18.00 | 2.0 | 12 | 12.8 | 71.6 | 46.6 | 64.2 | 21.3 | 22.6 |
| 1.00 | 1.50 | 21.00 | 3.0 | 6 | 0.4 | 56.9 | 0.4 | 75.7 | 7.8 | 1.3 |
| 1.00 | 1.50 | 20.00 | 3.0 | 6 | 0.5 | 62.5 | 0.4 | 75.7 | 7.4 | 1.3 |
| 1.00 | 1.50 | 19.00 | 3.0 | 6 | 0.4 | 62.0 | 0.4 | 75.7 | 7.3 | 1.3 |
| 1.00 | 1.50 | 18.75 | 3.0 | 6 | 0.4 | 58.5 | 0.4 | 75.7 | 7.3 | 1.3 |
| 1.00 | 1.50 | 18.50 | 3.0 | 5 | 0.2 | 67.3 | 0.0 | 79.1 | 1.7 | 1.7 |
| 1.00 | 1.50 | 18.25 | 3.0 | 4 | 0.1 | 95.1 | 0.0 | 87.1 | 0.5 | 0.4 |
| 1.00 | 1.50 | 18.00 | 3.0 | 4 | 0.2 | 94.7 | 0.0 | 87.1 | 0.6 | 0.4 |

# 8 Cross-identifications of correlated sources

## 8.1 Close X-ray clusters

There is yet no catalog of X-ray bright clusters covering the whole sky. The *Extended Medium Sensitivity Survey* of the EINSTEIN observatory (hereafter EMSS) covers only $\simeq 780$ square degrees, i.e., $\simeq 1.9\%$ of the whole sky. It is therefore not yet possible to decide which cases of strong correlation between a background source and X-ray emission are due to an X-ray luminous cluster. Nevertheless, there are two 1-Jansky sources having a correlation error level of less than 10% in all three photon-energy bands which ly within $60'$ of an EMSS cluster. One of them, Jy2216 − 038, was already found in BS3 to be strongly correlated with Lick galaxies. Table 6 contains the relevant data of the two sources and their foreground clusters.

**Table 7.** Strongly correlated 1-Jansky sources ($\epsilon < 10\%$ in all energy bands) with an EMSS cluster in the foreground ($z_{\rm cluster} < z_{\rm source}$) which is less than $1°$ away from the source. $\Delta\theta$ is the angular separation between source and cluster center in arc minutes, $L_{\rm X}$ is the (Einstein-de Sitter) X-ray luminosity of the cluster in $10^{44}\ h^{-2}{\rm erg/s}$. The clusters close to Jy2216 − 038 are seen in panel (b) of Fig.2

| Source | Cluster | $z_{\rm source}$ | $z_{\rm cluster}$ | $\Delta\theta$ (') | $L_{\rm X}$ $10^{44}\ h^{-2}{\rm erg/s}$ |
|---|---|---|---|---|---|
| Jy1127-145 | MS1127.7-141 | 1.19 | 0.11 | 46.2 | 3.89 |
| Jy2216-038 | MS2215.7-040 | 0.90 | 0.09 | 21.3 | 1.15 |
|  | MS2216.0-040 | 0.90 | 0.09 | 48.6 | 1.58 |

## 8.2 Sources correlated with galaxies *and* X-ray emission

The correlation analyses presented in BS2,3 have focussed on associations of foreground galaxies with background QSOs; the present study investigates associations of diffuse X-ray emission with the same sample of background objects. It was shown above that X-ray emission by discrete sources tends to introduce anticorrelations rather than correlations, which further indicates that we, in cases where correlations are found, deal with emission from extended sources. Galaxy clusters are the most likely objects in this respect. If the X-ray emission from galaxy clusters causes at least part of the correlations found here, there should be 1-Jansky sources which are strongly correlated with galaxies *and* X-ray emission. This is indeed the case, as summarized in Tab.8.

The table lists 1-Jansky sources which are strongly correlated (here, $r_{\rm corr} \geq 0.2$) with both Lick galaxies and X-ray emission (left part of the table) and with both IRAS galaxies and X-ray emission (right part). The table also distinguishes between the three photon-energy bands.

First of all, the number of multiple correlations decreases with increasing photon energy. This is at least partially due to the decreasing photon number density with increasing photon energy, causing larger relative random fluctuations in the photon counts per cell, and therefore also in the count ranks, and thus yielding less significant correlation results. The 1-Jansky sources in the table are probably not physically associated with the clusters potentially responsible for the correlations with both X-ray emission and galaxy counts, because most of them are at redshifts where clusters are unlikely to be detected in the *All-Sky Survey* and in the galaxy surveys.

**Table 8.** 1-Jansky sources which are highly correlated with X-ray emission *and* either Lick or IRAS galaxies. $m_{\rm QSO}$ is the visual magnitude of the source, $z_{\rm QSO}$ its redshift, and $r_{\rm corr}^{\rm Xray,Lick,IRAS}$ are the correlation coefficients with the respective foreground source types

| | Lick – X-ray | | | | | X-ray – IRAS | | | | |
|---|---|---|---|---|---|---|---|---|---|---|
| $E_\gamma$/keV | source | $m_{\rm QSO}$ | $z_{\rm QSO}$ | $r_{\rm corr}^{\rm Lick}$ | $r_{\rm corr}^{\rm Xray}$ | source | $m_{\rm QSO}$ | $z_{\rm QSO}$ | $r_{\rm corr}^{\rm Xray}$ | $r_{\rm corr}^{\rm IRAS}$ |
| $\geq 0.75$ | $2216-038$ | 17.00 | 0.90 | 0.59 | 0.70 | $1502+106$ | 18.60 | 0.56 | 0.75 | 0.45 |
| | $2134+004$ | 16.80 | 1.94 | 0.53 | 0.36 | $1055+018$ | 18.30 | 0.89 | 0.49 | 0.24 |
| | $1524-136$ | 21.00 | 1.69 | 0.35 | 0.33 | $2230+114$ | 17.30 | 1.04 | 0.46 | 0.27 |
| | $2136+141$ | 18.50 | 2.43 | 0.32 | 0.21 | $2155-152$ | 18.00 | 0.67 | 0.40 | 0.21 |
| | $2230+114$ | 17.30 | 1.04 | 0.30 | 0.46 | $1453-109$ | 17.40 | 0.94 | 0.39 | 0.21 |
| | $2223+210$ | 18.20 | 1.95 | 0.24 | 0.44 | $2149-306$ | 18.40 | 2.34 | 0.35 | 0.34 |
| | $1253-055$ | 17.80 | 0.54 | 0.23 | 0.33 | $1328+254$ | 17.70 | 1.05 | 0.29 | 0.22 |
| | $1005+077$ | 21.00 | 0.88 | 0.22 | 0.37 | $0834-201$ | 19.40 | 2.75 | 0.26 | 0.33 |
| | $1213+350$ | 20.10 | 0.85 | 0.22 | 0.39 | $1243-072$ | 18.00 | 1.29 | 0.26 | 0.32 |
| | $1502+106$ | 18.60 | 0.56 | 0.21 | 0.75 | $0332-403$ | 18.50 | 1.45 | 0.25 | 0.20 |
| $\geq 1.00$ | $2216-038$ | 17.00 | 0.90 | 0.59 | 0.72 | $1502+106$ | 18.60 | 0.56 | 0.58 | 0.45 |
| | $1524-136$ | 21.00 | 1.69 | 0.35 | 0.25 | $2155-152$ | 18.00 | 0.67 | 0.35 | 0.21 |
| | $0420-014$ | 17.70 | 0.92 | 0.28 | 0.34 | $1453-109$ | 17.40 | 0.94 | 0.35 | 0.21 |
| | $2223+210$ | 18.20 | 1.95 | 0.24 | 0.28 | $0420-014$ | 17.70 | 0.92 | 0.34 | 0.26 |
| | $1005+077$ | 21.00 | 0.88 | 0.22 | 0.43 | $0332-403$ | 18.50 | 1.45 | 0.28 | 0.20 |
| | $1213+350$ | 20.10 | 0.85 | 0.22 | 0.42 | $1055+018$ | 18.30 | 0.89 | 0.27 | 0.24 |
| | $1502+106$ | 18.60 | 0.56 | 0.21 | 0.58 | $1243-072$ | 18.00 | 1.29 | 0.23 | 0.32 |
| | | | | | | $0804+499$ | 17.50 | 1.43 | 0.20 | 0.22 |
| $\geq 1.25$ | $2216-038$ | 17.00 | 0.90 | 0.59 | 0.38 | $0804+499$ | 17.50 | 1.43 | 0.37 | 0.22 |
| | $2134+004$ | 16.80 | 1.94 | 0.53 | 0.54 | $2155-152$ | 18.00 | 0.67 | 0.35 | 0.21 |
| | $1524-136$ | 21.00 | 1.69 | 0.35 | 0.35 | $1502+106$ | 18.60 | 0.56 | 0.31 | 0.45 |
| | $2223+210$ | 18.20 | 1.95 | 0.24 | 0.24 | $1055+018$ | 18.30 | 0.89 | 0.29 | 0.24 |
| | $1049+215$ | 18.50 | 1.30 | 0.21 | 0.27 | | | | | |
| | $1502+106$ | 18.60 | 0.56 | 0.21 | 0.31 | | | | | |

# 9 Summary and discussion

## 9.1 Data and results

We have investigated whether diffuse X-ray emission in the energy range $E_\gamma \in [0.75, 2.2]$ keV is correlated with 1-Jansky sources above redshift 0.5 on scales of $\gtrsim 10'$. For that purpose, we have applied rank-order statistics as described in BS1,2,3 to fields taken from the ROSAT *All-Sky Survey*. Of these fields, 246 were centered on the positions of the 1-Jansky sources, and 246 were chosen such that they had equal galactic latitude as the source fields, but random galactic longitude. Subsamples of source fields were selected by the redshifts and the optical (visual) magnitudes of the sources, to see whether the results depend on these parameters. Each control field was formally assigned the redshift and the optical magnitude of the source field with the same galactic latitude, so that the choices of redshift- and magnitude intervals equally changed the number of fields in the control sample. The 'strength' of correlations is quantified via the rank-order correlation coefficient $r_{\mathrm{corr}}$, and, equivalently, the correlation error level $\epsilon$.

The results presented in Figs.3,4 of Sect.4 and in Tabs.1,2 show that

(1) there is evidence at a significance level of up to 99.8% for correlations between distant ($z \gtrsim 1.5$), optically bright ($m \lesssim 19$) 1-Jansky sources and diffuse X-ray emission quite irrespective of the X-ray photon energy $E_\gamma$;

(2) there is evidence on a slightly less significance level ($\leq 97.3\%$) for correlations between 'close' ($z \lesssim 1.0$) 1-Jansky sources and diffuse X-ray emission with $E_\gamma \gtrsim 0.75$ keV;

(3) in the intervening source redshift interval $1.0 \lesssim z \lesssim 1.5$, there is no sign of significant correlations, irrespective of the source-subsample parameters and the X-ray photon energy; however, when the radio flux threshold is increased, highly significant correlations are obtained (cf. Tab.6);

(4) the results obtained from source fields and those obtained from control fields are clearly distinct, to be seen most clearly in Fig.4;

(5) the control fields tend to show *anti*correlations with X-ray emission, which reach a significance of up to 100%.

Although on a weaker significance level, the Kolmogorov-Smirnov tests performed in Sect.5 support the results listed above. Also, the results from bootstrapping source- and control fields indicate that the correlation- or anticorrelation effects are *not* due to a few sources, because a bootstrapped sample misses on average $1/\mathrm{e} \simeq 37\%$ of the original members.

Item (5) of the above list, disturbing at first sight, finds a simple explanation in Sect.6, where we show with the help of Monte-Carlo simulations that these anticorrelations are due to discrete sources whose emission is superposed on the otherwise diffuse X-ray background. This explanation also means that the correlations found in the source-field subsamples actually are more significant than given in the left columns of Figs.3,4 and Tab.1, because the correlations have to 'come up against' the anticorrelations which are naturally present because of the discrete-source contamination of the fields. It is, however, hardly possible to quantify this effect because it depends on the extension of the contaminating sources, which in turn depends on their brightness and their hardness.

The reason for item (3) of the above list, i.e., the existence of a source-redshift interval where no significant correlations are found, is unclear to us. Again, the bootstrapping results indicate that the lack of significant correlations in this redshift band is not caused by a few sources only. Possibly, the correlations found for 'low' source redshifts ($z \lesssim 1.0$) are caused by different sources of diffuse X-ray emission than in the case of distant sources ($z \gtrsim 1.5$). Some support for such a hypothesis may be that the distant sources are correlated with soft and hard X-ray photons, while the close sources are correlated with soft X-ray emission and show no significant correlation with hard X-ray emission. If, however, the correlations for close sources were due to possible host clusters of the sources themselves, these clusters would be at redshifts $0.5 \lesssim z \lesssim 1.0$, and the maximum of their X-ray luminosity should be around $\simeq 2\ldots 3$ keV, i.e., in they should be brightest in the 'hard' photon band of our study ($E_\gamma \in [1.5, 2.2]$ keV), while we see the correlations in this redshift range only for softer photons. In the 'hard' band, however, the photon number density is lowest and therefore most affected by Poisson noise.

The curves in panels (III) and (IV) of Fig.8d drop to zero at $\simeq 15'$, an angular scale roughly corresponding to the Abell radius of galaxy clusters at intermediate redshifts. Closer inspection reveals that this increase towards the field center is primarily due to sources in the redshift regime $[0.5\ldots1.0]$, while it is less pronounced and off-center for higher-redshift sources. This could mean that the enhanced X-ray emission around 1-Jansky sources is due to the emission of possible QSO host clusters. If, as we will argue below, the X-ray emission around high-redshift 1-Jansky sources is due to gravitationally lensing galaxy clusters in their foreground, these clusters would not be expected to be centered on the background sources, and they would also appear less extended because of their larger average distance and their weaker average flux.

## 9.2 Interpretations and prejudices

It can definitely be said on the basis of the above that the positions of 1-Jansky sources do not resemble *randomly* chosen positions on the X-ray sky as seen by ROSAT in the *All-Sky Survey*: the control fields, which were randomly selected, show a distinctly different behaviour than the source fields. We have taken care to cut out the X-ray emission of the 1-Jansky sources themselves by omitting the central circle of 100 square arc minutes of each field, which should definitely be safe in view of the PSPC point-spread function. The correlation scale is on the order of $10'\ldots20'$, corresponding to $\lesssim 5/h$ Mpc at intermediate redshifts, $h$ being the Hubble constant in units of 100 km/s/Mpc. This is a scale typical for galaxy clusters, and at the same time clusters are the largest contingent X-ray emitters known in the Universe.

If the sky positions of 1-Jansky sources are distinguished from random positions by being close to galaxy clusters, two possibilities arise: either the 1-Jansky sources are physically associated with the clusters, or they are gravitationally lensed by them. Let us assume for the moment that the detected correlations are due to the possible host clusters of the 1-Jansky sources. Then, the clusters have to have the same redshifts as the 1-Jansky sources. From the excess counts in the central region of the source fields (see Fig.8d, panel (II)) and $\langle n_\gamma \rangle \simeq 0.2$ arcmin$^{-2}$, we estimate an excess count rate from the hypothetical clusters of $10^{-2}$ s$^{-1}$. Assuming a thermal spectrum of $10^7$ K and a mean galactic hydrogen column density of $3\times10^{20}$ cm$^{-2}$, we estimate an average luminosity of $8\times10^{43}\ h^{-2}$erg/s for such clusters at a redshift of 0.5. If, e.g., $\simeq 10\%$ of all 1-Jy sources have a possible host cluster, they do not need to be extremely luminous to be statistically detected in the *All-Sky Survey*, although they would be much too faint to be individually detected. This means that for $0.5 \lesssim z \lesssim 1.0$ we might indeed see a possible host-cluster emission, with the tendency that optically brighter 1-Jy sources have a more luminous host. For higher redshifts, however, the clusters would have to exceed a luminosity of $10^{45}\ h^{-2}$erg/s to be detected. We therefore conclude that possible host clusters may indeed cause the correlations seen for the lowest source redshifts, but that this is very improbable for the highest redshifts, since then the clusters would have to be extremely luminous, $L_X \gtrsim 10^{45\ldots46}\ h^{-2}$erg/s.

On the other hand, it is striking to see in Tab.8 that there is a number of highly correlated 1-Jansky sources which are correlated not only with X-ray emission, but also with either Lick or IRAS galaxies. If the X-ray emission is due to clusters, the correlated galaxies should belong to these clusters. The galaxies in the Lick or IRAS catalogs, however, are certainly in the foreground of the 1-Jansky sources, because the catalogs are not deeper than $z \lesssim 0.2$ for Lick and $z \lesssim 0.4$ for IRAS (for the Lick catalog, see Shane & Wirtanen 1967, Seldner et al. 1977, for the IRAS *Faint Source Catalog*, cf.

Strauss et al. 1990, Yahil et al. 1991, and for the redshift distribution of IRAS galaxies, see Saunders et al. 1990). This indicates that at least for those sources which appear in Tab.4, if the X-ray emission comes from clusters, they should be in the foreground, which then can hardly avoid lensing the background sources, thereby magnifying them, and causing a magnification bias. The sources listed in Tab.8 are also natural targets to look for coherent distortions of the images of background galaxies (cf. Bonnet et al. 1993).

Further support for the lensing hypothesis is provided by the result (apparent in the left columns of Figs.3,4 and Tabs.1,2) that the correlations (1) are stronger for high than for low source redshifts and (2) that the correlations tend to strengthen when the source subsamples become optically brighter. If the radio emission of the 1-Jansky sources is at most weakly correlated with their optical emission, the multiple-waveband magnification bias effectively steepens the joint luminosity function of the sources and thereby enhances the magnification bias from lensing. The strengthening of the correlations with increasing optical source brightness therefore tentatively argues for the lensing effect. Note also in Tab.2 that strong correlations for optically bright, distant 1-Jansky sources appear despite the fact that the source subsamples are very small, indicating that most of the sources in such subsamples are highly correlated with X-ray emission and not just a few. For increasing source redshift, we expect the lensing frequency to rise, because the optical depth for magnification increases with increasing volume between the sources and the observer.

A final, admittedly conjectural argument for the X-ray correlation to come from the foreground and not from host clusters goes as follows. Observations of clusters show that clusters probably are young (e.g., Geller & Beers 1982, Gioia et al. 1990, Henry et al. 1992, Henry 1992), and it is also expected from theoretical arguments that the spatial number density of clusters evolves rapidly between $z \lesssim 0.7$ and today, if the density of the Universe is close to its critical (Einstein-de Sitter) value (Richstone et al. 1992, Bartelmann et al. 1993). This means clusters should be rare at redshifts $z \gtrsim 0.5$, and therefore also possible host clusters of 1-Jansky sources should be a negligible phenomenon for distant sources. Moreover, the evolution of the X-ray luminosity function of clusters shows that the spatial number density of X-ray luminous clusters decreases with increasing redshift, making it improbable that the correlations of high-redshift 1-Jy sources with X-ray emission is due to clusters physically associated with them.

In any case, Tab.7 shows that there are at least two sources in our sample, $Jy2216-038$ and $Jy1127-145$, which are 'close' to known X-ray luminous foreground clusters. If the clusters can be modelled by singular isothermal spheres with a velocity dispersion of $\simeq 1200$ km/s, they are expected to magnify the sources by $\lesssim 10\%$ because of the large distance between the cluster centers from the sources.

*Acknowledgements*. We are grateful to H. Fink and J. Trümper, who made this study possible. For intense support with the EXSAS software package, we wish to thank H.-C. Thomas, and, in particular, M. Freyberg, whose comments and hints were of invaluable help to us. The EINSTEIN *Medium Sensitivity Survey* was provided by the Einstein On-line Service, Smithsonian Astrophysical Observatory. Last not least, C. Rosso is acknowledged for her patience while retrieving the data of all the photons in the *All-Sky Survey* detected in nearly 2000 square degrees of the sky.